\documentclass[fleqn,usenatbib]{mnras}

\usepackage{newtxtext,newtxmath}

\usepackage[T1]{fontenc}
\usepackage{ae,aecompl}

 \usepackage{times}
 \usepackage{rotating}
\usepackage{lscape}
 \usepackage{natbib,twoopt}
 \usepackage[normalem]{ulem}

\usepackage{longtable}

\usepackage{graphicx}	
\usepackage{amsmath}	
\usepackage{amssymb}	

\usepackage{enumitem}

\title[The radius inflation in short-period low-mass binaries]{The radius inflation problem in short-period low-mass binaries: a large sample analysis}

\author[H. E. Garrido et al.]{
Hern\'an E. Garrido$^{1,2,3}$\thanks{E-mail: hgarridov@ecci.edu.co},
Patricia Cruz$^{3}$, Marcos P. Diaz$^{3}$,
John F. Aguilar$^{4}$\\
$^{1}$Grupo de Simulaci\'on, An\'alisis y Modelado, Vicerrector\'ia de Investigaci\'on, Universidad ECCI, Cra. 19 No. 49-20, Bogot\'a, Colombia\\
$^{2}$Observatorio Astron\'omico, Universidad Distrital Francisco Jos\'e de Caldas, Calle 13 No.31-75, Bogot\'a, Colombia\\
$^{3}$Instituto de Astronomia, Geof\'isica  e Ci{\^e}ncias Atmosf\'ericas, Universidade de S{\~a}o Paulo, Rua do Mat{\~a}o 1226, Cidade Universit\'aria,\\ 05508-090 S{\~a}o Paulo, Brazil\\
$^{4}$Departamento de Matem\'aticas, Universidad Militar Nueva Granada, kil\'ometro 2 v\'ia Cajic\'a - Zipaquir\'a, Colombia, c\'odigo postal 110111.}

\date{Accepted 2018 November 01. Received 2018 November 01; in original form 2018 April 04}

\pubyear{2018}

\begin{document}
\label{firstpage}
\pagerange{\pageref{firstpage}--\pageref{lastpage}}
\maketitle

\begin{abstract}
Due to the recent increase in the availability of photometric time-series databases, 
the characterisation of low-mass eclipsing binaries for the study of their orbital and physical parameters is now possible in large samples and with good precision.
We have identified and photometrically characterised a sample of 230 detached close-orbiting eclipsing binaries with low-mass main-sequence components in the Catalina Sky Survey. These low-mass stars have masses of $M \leq 1.0\ M_{\odot}$ and orbital periods shorter than $2$ days. The adopted method provided a robust estimate of stellar parameters (as mass and fractional radius) by using only light curves and photometric colours, since no spectroscopic information was available for these objects. 
A SDSS-2MASS ten-colour grid of composite synthetic and observed colours and the K-Nearest Neighbours method were employed to identify main-sequence stars and to estimate their effective temperatures, typically of $T_{\rm eff}\leq 5720$ K. 
Each light curve was modelled with the JKTEBOP code together with an asexual genetic algorithm to obtain the most coherent values for the fitted parameters. The present work provides an unprecedented number of homogeneous estimates of main stellar parameters in short-period low-mass binary systems. The distribution of the components of the investigated detached eclipsing binaries in the mass-radius diagram supports a trend of radius inflation on low-mass main-sequence stars. A relative increase of inflation for lower masses is also found and our results suggest that the secondaries are more inflated, i.e. they present larger radii than the primary components of same mass, when compared to stellar evolutionary models. 

\end{abstract}




\section{Introduction}\label{intro}

Stellar evolutionary models are widely used to estimate fundamental properties of stars and stellar populations. However, when compared to observations of masses and radii derived with small uncertainties -- of less than 5\%, for instance --, the theoretical stellar models yield radius values up to 10\% smaller than measurements for low-mass stars \citep[][and references therein]{Kraus2011,Birkby2012,Dittmann2017}. This discrepancy is known as the radius anomaly of low-mass main-sequence stars in binary systems and it is still an unsolved problem of stellar astrophysics.

A few explanations for the observed radius inflation have been discussed in the literature, where the radius anomaly have been associated with non-solar metallicity \citep{Berger2006, Lopez-Morales2007}, or pointed to be caused by an enhanced magnetic activity \citep{Chabrier2007,Kraus2011}, or even be due to variations in the light curve caused by stellar spots or flares \citep{Morales2008, Morales2010}. For example, \citet{Berger2006,Lopez-Morales2007} have found evidence that relates isolated inactive M stars with inflated radii to stars with higher metallicities. Alternatively, \citet{Chabrier2007} and \citet{Kraus2011} suggested that M dwarfs as components of close detached binary systems, with short orbital periods of just a few days, should have their magnetic activity enhanced by tidal effects, which would reduce the convective efficiency and lead to an inflation of the stellar radius.

Eclipsing binary (EB) systems are excellent targets for the determination of stellar fundamental properties and provide the best opportunity for testing stellar evolutionary theories. Moreover, detached eclipsing binaries (DEBs) provide the most accurate method for obtaining stellar masses and radii \citep{Andersen1991,Coughlin2011}. However, until now only a small number of DEBs with late-K or M-type components have been characterised with good precision \citep[][and references therein]{Birkby2012,Cruz2018}. 
For instance, \citet{Southworth2015} maintain an on-line catalogue of detached EB systems with mass and radius derived with high accuracy (of $\sim$$2$\%), with around 40 systems where the most massive component (primary) has a mass of $M\leq1$ $M_{\rm \odot}$. If only close systems are considered, with orbital period of less than 2 days, this number is smaller, less 20 short-period DEBs are well characterised. \citet{Eker2014} also have published a catalogue of detached double-lined EBs, but gathering from the literature systems with physical parameters determined with larger uncertainties. Among them, over 60 DEBs have primaries with mass of $M\leq1$ $M_{\rm \odot}$, but only 35 of them have periods shorter than 2 days.

Given the availability of extensive photometric time-series databases, the characterisation of low-mass eclipsing binary systems (LMEBs) has become possible in large samples, where trends can be seen despite the large individual uncertainties. Deriving masses and radii in statistically significant samples may allow the investigation of the correlation between radius inflation and other basic stellar parameters.

Knowing that the number of known detached short-period systems in the literature is limited, this work is then dedicated to identify and photometrically characterise short-period detached eclipsing binaries, with low-mass main-sequence stars (LMMS) as components, in the Catalina Sky Survey \citep{Drake2009}, and to increase the number of such systems --DEBs with $P_{0}<2$ d and $M\leq1.0$ $M_{\rm \odot}$-- with studied orbital configuration and derived stellar parameters.

We firstly describe the sample selection in Sect. \ref{select}, followed by the spectral characterisation of the individual stellar components in Sect. \ref{Spectral-types}. The light curve modelling procedures used to characterise the DEB systems are presented in Sect. \ref{modelling}, along with the results. Finally, Sect. \ref{discuss} and \ref{concl} are dedicated to the discussion of the derived mass-radius distribution and conclusions, respectively.

\section{The Sample Selection}\label{select}

The Catalina Sky Survey \citep[CSS, ][]{Drake2009}, is a project dedicated to search and catalogue rapidly moving near-Earth objects (NEOs), and those large NEOs that can possibly become an impact threat to Earth\footnote{For more details, see \url{https://catalina.lpl.arizona.edu}}. This project uses three telescopes (1.5, 1.0, and 0.75 meter-telescopes) and covers objects in a wide portion of the sky (from declination $\delta = -75^\circ$ to $\delta = +70^\circ$).

\citet{Drake2014} presented an on-line catalogue with $47000$ periodic variable stars identified within the Catalina Surveys Data Release-1 \citep[CSDR1, ][]{Drake2012}. The published catalogue contains several quantities such as the Catalina V magnitude ($V_{\rm CSS}$), the identified period and amplitude, and a designed classification for each variable candidate. 
To identify and select DEB systems with low-mass main-sequence components --defined in this work as stars with mass of $1.0$ $M_{\odot}$ or less--, we performed the steps described bellow.

Firstly, we concentrated our search on binaries preliminary assigned as {\sl Algol type} (EAs) in the above mentioned catalogue, as well as those designed as {\sl Class 2} objects. We then selected those objects with an identified orbital period of less than $2$ days. At this point, we have found 3946 short-period DEB candidates. 
We then obtained the complete light curves (LCs) for all selected candidates from the Catalina Surveys Data Release-2\footnote{\url{http://nesssi.cacr.caltech.edu/DataRelease/}} (CSDR2), which comprehends all seven years of photometric data. All candidates that presented less than 250 individual measurements (N$_{\mathrm{obs}}<250$) in their light curves were discarded. Each light curve was visually inspected in order to verify if they correspond to detached systems and to minimise the amount of semi-detached binaries or with highly significant reflection effects in the selected sample. We also performed a $3\sigma$ clipping on the data for the removal of outliers.

The light curves from the sample were pre-analysed to obtain a few orbital parameters, which were refined later during the analysis described in Sect. \ref{modelling}. We have defined as primary component the star being occulted during the deepest eclipse -- i.e. the hottest and the most massive star in a V+V system --, and thus we derived the reference time of the primary minimum ($T_{0}$) as the time of the primary mid-eclipse. The orbital period was then obtained by using the phase dispersion minimisation algorithm introduced by \citet{Stellingwerf1978}, which is implemented in IRAF\footnote{IRAF is distributed by the National Optical Astronomy Observatory, operated by the Association of Universities for Research in Astronomy, Inc., under cooperative agreement with the National Science Foundation.}. Only those candidates with periods of less than $2$ days were kept in the sample.

Within the remaining candidates, the next selection criterion was the availability of additional photometry from point-source catalogues between $0.3-2.1\,\mu\mathrm{m}$, namely the Sloan Digital Sky Survey \citep[SDSS, ][]{Abazajian2009} and the Two Micron All Sky Survey \citep[2MASS, ][]{Skrutskie2006}. For that, we performed a conservative search within a radius of $1.5''$ and selected the closest match for each candidate by using TOPCAT\footnote{\url{http://www.star.bris.ac.uk/~mbt/topcat/}} \citep[Tool for OPerations on catalogueues And Tables, ][]{Taylor2011}. This search has reduced the number of DEB candidates to 534 objects, all of them with additional photometry in eight broad bands (SDSS $ugriz$ and 2MASS $JHK$).

 \begin{figure}
\includegraphics[width=0.475\textwidth]{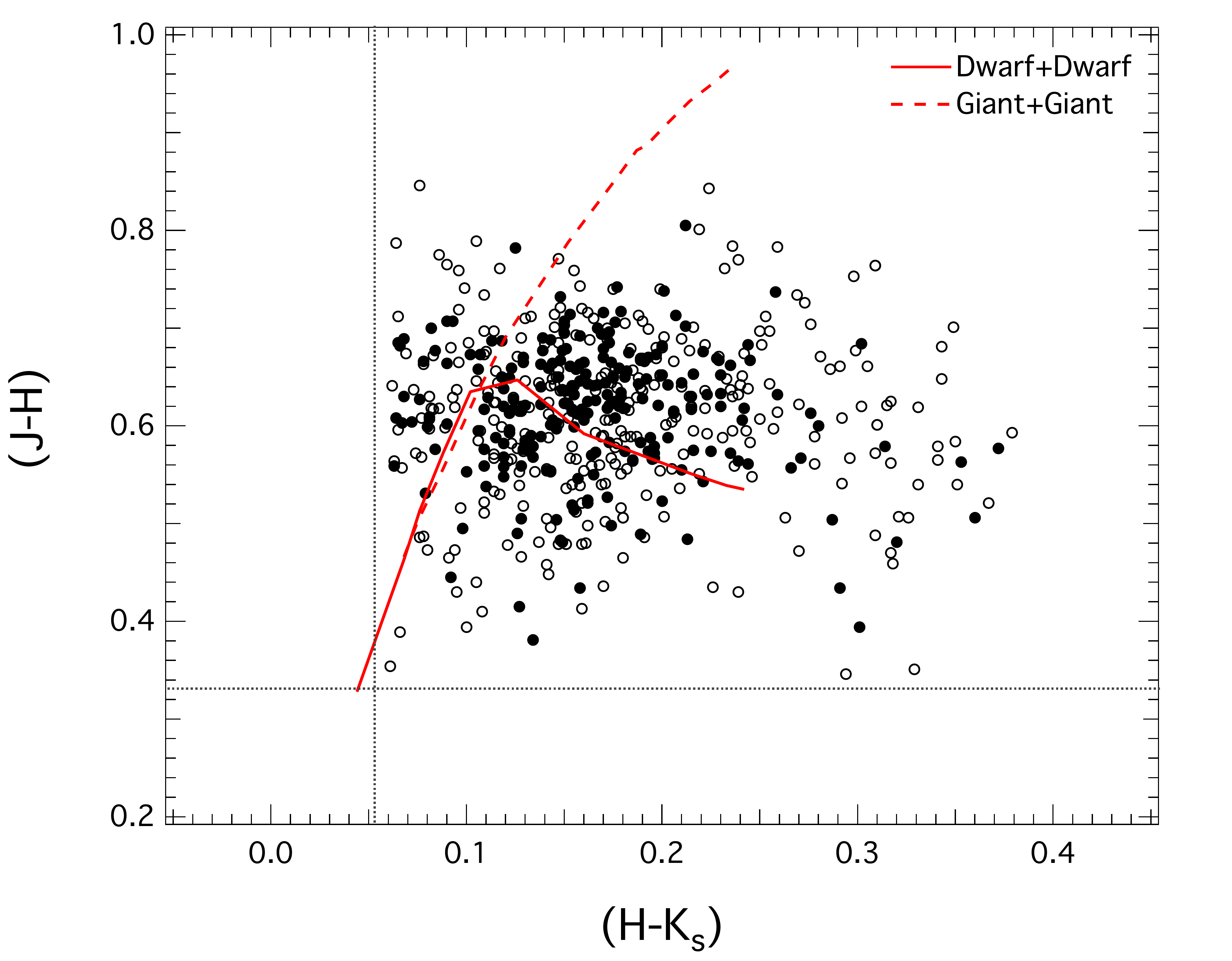}
\caption{colour-colour diagram of 534 pre-selected objects from the Catalina Sky Survey, represented by open circles. These candidates present light curves containing more than 250 individual measurements (N$_{\mathrm{obs}}\geq 250$) and additional photometry from SDSS and 2MASS catalogues. The filled circles represent a selection of 230 DEB candidates that have composite colours compatible with two main-sequence stars, with luminosity class V+V, obtained after applying the KNN classifier described in the Sect. \ref{Spectral-types}. The sequences for combinations of luminosity classes V+V and III+III, obtained by using the method described in Sect. \ref{Spectral-types} are shown in red solid and dashed lines, respectively. The dotted lines show the colours corresponding to a G3-dwarf star ($T_{\mathrm{eff}}=5720$ K, $M=1.0\, M_{\odot}$)}
    \label{NIR-CMD-534stars}
\end{figure}

The next step was to use the available photometry to derive candidate's colours for the characterisation of each DEB component. Starting with 2MASS and Catalina magnitudes, $JHK$ and $V_{\rm CSS}$ respectively, we obtained the $J-H$, $H-K$ and $V-K$ colours. We then constructed the ($H-K$) $-$ ($J-H$) diagram with our DEB candidates to identify the population of stars with mass $M\lesssim 1.0\,M_{\odot}$, as shown in Fig. \ref{NIR-CMD-534stars}. The dotted lines show the colours corresponding to a G3-dwarf star, with an effective temperature of $T_{\mathrm{eff}}=5720$ K and mass of $1.0\,M_{\odot}$, where our selection cut was placed, thus all 534 pre-selected candidates correspond to stars with $M\leq 1.0\,M_{\odot}$ and $T_{\mathrm{eff}}\leq 5720$ K.

It is important to take into account that our candidates are binary systems and, therefore, the observed colours are the result of the contribution of two components with different absolute magnitudes, each one weighted by a corresponding luminosity. Then, the composite colours can be the result from three different combinations of luminosity classes: V+V, V+III or III+III. Hence, the final criterion for our sample selection is to discard binary systems with a giant star as component, whose composite colour results from the combinations of components with luminosity classes V+III and III+III.

\citet{Bessell1988} showed the position of dwarfs and giants schematically in the ($H-K$) $-$ ($J-H$) diagram and that these two different luminosity classes share similar ($J-H$) and ($H-K$) colours for ($V-K$) $< 3.5$. Hence, the separability of such systems become uncertain when considering only these colours. 
Therefore, to identify only V+V systems, we adopted an independent method based on a total of ten colour indices, combining the photometry from SDSS and 2MASS, which is described in detail in the following section (Sect. \ref{Spectral-types}). In figure \ref{NIR-CMD-534stars}, the sequences expected for binary systems composed by III+III and V+V stars, with the contribution from both components combined, were obtained by this method and are represented by red dashed and solid lines, respectively, for comparison. In the end, our final sample is composed by 230 short-period DEB candidates with low-mass main-sequence components.

 \begin{figure*}
    \centering
\includegraphics[width=0.75\textwidth]{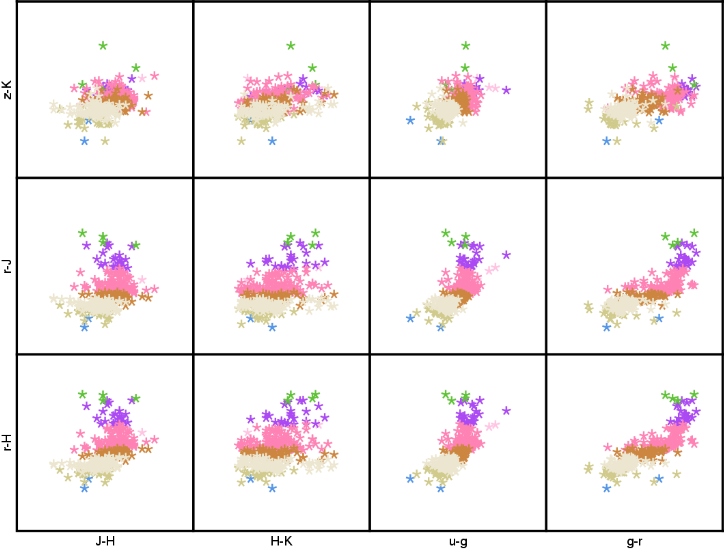}
\caption{
Examples of composite SDSS and 2MASS colours of all DEB candidates, distributed in 8 clusters by the KNN method. These candidates are those systems identified as a combination of two main-sequence stars (V+V) and that have temperatures assigned by our method described in sect. \ref{Spectral-types}. The colours represent different clusters, which are grouped by similarity to the synthetic data.}
    \label{obs_colourgrid}
\end{figure*}

\begin{table*}
\centering
\scalebox{0.8}[0.9]{
\begin{tabular}{l |cccc|cccc|c}
\hline	
   && \multicolumn{2}{|c|}{This work}   & &\multicolumn{4}{|c|}{Literature}\\
Star Name        &	$\mathrm{T}_1$ (K)          	&	 $\mathrm{T}_2$ (K)        &  $\mathrm{M}_1$ (M$_{\odot}$) &  $\mathrm{M}_2$ (M$_{\odot}$)   &   $\mathrm{T}_{1,pub}$ (K)       &	 $\mathrm{T}_{2,pub}$ (K)	      &	 $\mathrm{M}_{1,pub}$ (M$_{\odot}$)		    &		$\mathrm{M}_{2,pub}$ (M$_{\odot}$)        & Reference \\
\hline
SDSS-MEB-1  & $3428\pm 100$   &	$3254\pm 150$ &	$ 0.404\pm 0.058 $ & $0.266 \pm 0.166$	& $3320\pm130$  &	$3300\pm130$  &	$0.272\pm 0.020$    &	$0.240\pm 0.022$  & \cite{Blake2008} \\
GU Boo 	    &	$4302\pm 164$   &	$3428\pm 100$& $0.708\pm 0.013$  & $0.404\pm 0.058$ & $3920\pm130$ &	$3810\pm 130$ &	$0.610\pm 0.007$    &	$0.599\pm 0.006$  & \cite{Windmiller2010}\\
19b-2-01387 &	$3392\pm 115$   &	$3228\pm 107$ & $0.383\pm 0.062$  &	 $0.244\pm 0.091$  & $3498\pm 100$     &	$3436\pm 100$   &  $0.498\pm0.019$     &	$0.481\pm 0.017$ & \cite{Birkby2012} \\	
MG1-646680  & $3663\pm 100$   &	$3407\pm 166$ & $0.508\pm 0.052$  &	$0.394\pm 0.091$  & $3730\pm 20$  &	$3630\pm 20 $&	$0.499\pm0.002 $    &	$0.443\pm0.002 $ & \cite{Kraus2011}\\ 		
19g-4-02069    &	$3428\pm 100$   &	$3254\pm 150$     & $0.404\pm 0.058$    &  $0.266\pm 0.124 $    &	$3300\pm 140$ &	$2950\pm 140$ &	$0.53\pm 0.02$	    &	$0.143\pm 0.006$  & \cite{Nefs2013} \\
CU Cnc	       &	$3150\pm 100$   &	$3150\pm 100$   &  $0.200\pm 0.063$   &  $0.200\pm 0.063$   &   $3160\pm 150$    &	$3125\pm 150$    &	$0.4333\pm 0.0017$  &	$0.3980\pm0.0014$ & \cite{Ribas2003}\\
19c-3-01405    &	$3288\pm 120$   & 	$3150\pm 100$      & $0.295\pm 0.085$   & $0.200\pm 0.063$   &	$3309\pm130$  &	$3305\pm139$  &	$0.410\pm0.023$    &	$0.376\pm0.024$  &  \cite{Birkby2012}\\
 LSPM J1112+7626  &  $3575 \pm100 $ & $3387 \pm 179$ & $0.426 \pm 0.040 $ & $0.351 \pm 0.072 $ & $ 3191\pm 164$ & $3079 \pm 166$ & $ 0.395\pm 0.002$	& $ 0.275\pm 0.001$ &	\cite{Irwin2011}\\
 HAT-TR-318-007   &  $ 3428\pm100 $ & $3254 \pm 146$ & $ 0.367\pm 0.040$ & $ 0.297\pm 0.059 $ & $ 3190\pm 110$ & $ 3100\pm 110$ & $ 0.448\pm 0.011$	& $ 0.2721 \pm 0.0042$ &	\cite{Hartman2018}\\
  V1236 Tau      &  $ 4551\pm 709$ & $ 3612\pm 100$ & $ 0.835\pm0.142 $ & $ 0.441\pm 0.040$ &   $ 4200\pm 200$ & $ 4150\pm 200$ & $ 0.787\pm0.012 $ &	$0.770 \pm 0.009$ &	\cite{Bayless2006}\\
19c-3-08647		& $3761 \pm 100$ & $3491\pm204 $ & $0.500 \pm 0.011$ & $ 0.392\pm 0.082$ &	$ 3900\pm 100$	&	$ 3000\pm 150$	&	$0.393 \pm 0.019$	&	$ 0.244\pm 0.014$ &		\cite{Cruz2018}\\
19f-4-05194		& $4234 \pm 100$ & $3829 \pm 299$ & $0.771 \pm 0.020$ & $ 0.528\pm 0.120$ &	$ 4400\pm 100$&		$3500 \pm 100$	&	$ 0.531 \pm 0.016 $	&	$ 0.385\pm 0.011$	&	\cite{Cruz2018}\\
19g-2-08064		& 4085 $ \pm116 $ & $3761 \pm 261$ & $ 0.741\pm0.023 $ & $0.500 \pm 0.105$ &	$ 4200\pm 100$	&	$ 3100\pm 100$	&	$ 0.717 \pm 0.027$	&	$ 0.644\pm 0.025$	&	\cite{Cruz2018}\\  
 SAO 106989  &  $5935 \pm 100$ & $3686 \pm706 $ & $1.035 \pm 0.030$ & $ 0.470\pm 0.282$ &	      $6000\pm100$ &  $2380\pm 259$ & $1.11\pm 0.22$	&	$0.256 \pm 0.005$ &	\cite{Chaturvedi2018}\\
EPIC 211682657    &  $ 5935\pm 100$ & $ 3775\pm 532$ & $ 1.035\pm0.030 $ & $0.506 \pm 0.213 $ &   $6650\pm 150$ & $4329\pm 49$ & $1.721\pm 0.047$ &	$0.599 \pm 0.017$ &	\cite{Chaturvedi2018}\\
 \hline
\end{tabular}}
\caption{Determination of  effective temperatures and masses using the k-means method described in Sect. \ref{Spectral-types}, for a small control sample of low-mass binaries from the literature, with available 2MASS and SDSS colours. Columns 2-to-5 present the results obtained by our method and columns 6-to-9 show the values from the literature.} 
\label{Table-Literature-LMB}
\label{Table:1}
\end{table*}

\section{The effective temperature and mass of individual components}\label{Spectral-types}

We derive the effective temperature of the individual components by using the available multi-wavelength photometry from SDSS and 2MASS bands of our targets. The applied technique is based on the method used by \citet{Parihar2009}, and the implementation of a supervised statistical analysis that allows the classification of our multi-colour dataset as explained bellow. 
This method was used to select binaries composed only by main-sequence stars in our sample, with masses of $M \leq 1.0\, M_{\odot}$. For that aim, we have used different models with any combination of temperatures between $3150-5935$ K for main-sequence (V) stars and $3428-5050$ K for giant (III) stars, with masses in the interval of $0.10\, M_{\odot}\leq M \leq 2.15\, M_{\odot}$. However, it is important to state that the method is not intended to separate all systems into the three possible EB populations (V+V, V+III or III+III). The objective was to select EB systems with only dwarfs as components (V+V systems), by separating them from those with giant stars (V+III or III+III).

At first, the SDSS-2MASS ten-colour calibration grid of synthetic composite colours is constructed. We used the seven standard colours with adjacent passbands, namely (u-g), (g-r), (r-i), (i-z), (z-J), (j-H), and (H-K). We added three redder colours, (r-J) (r-H), and (z-K), to help the method to distinguish better V+V binaries from those with giant stars. These colours allow us to obtain a stellar classification in terms of effective temperatures and luminosity classes of the binary components, considering the colour indices that correspond to a combination of luminosity classes (V+V, V+III or III+III). A complete synthetic colour grid is created from the 990 possible temperature combinations using 30 different temperatures for main-sequence stars and another 14 for giants. We use the \citet{Bressan2012} 1 and 3 Gyr isochrone models with a metallicity of $Z = 0.015$ to generate the composite synthetic colours using the method described in \citet{Parihar2009}.

We then group the synthetic model colours and the observed colours from the 534 pre-selected objects in a single ten-colour matrix and implement a neighbours-based classification. The K-Nearest Neighbours (KNN) classifiers \citep{Hartigan1975,Chattopadhyay2014} are grouping algorithms that iterate to assign objects of a sample in a cluster of objects with similar characteristics, which have to be previously defined by some validation method \citep{Albalate2009, Telgarsky2010}. 
In this work, the Hartigan's method was adopted as the cluster validation technique for our multi-colour dataset. Initially proposed by \citep{Hartigan1975}, the following metric was described by \citet{Zhao2009} for detecting the optimum number of clusters, $k$, to be applied in the $k$-means algorithms:

\begin{equation}
H(k)=-\log \Bigg(\frac{SSW}{SSB}\Bigg),
\end{equation}
where
\begin{equation}
SSW (C,m)=\frac{1}{n}\sum_{i=1}^{m}\sum_{j \in \,C_{i}}^{}\Arrowvert x_j -C_{P(j)}\Arrowvert
\end{equation} 
is the quadratic sum between clusters and 
\begin{equation}
SSB(C,m)=\frac{1}{n}\sum_{i=1}^{m} n_{i}\Arrowvert C_i - \bar{x} \Arrowvert 
\end{equation} 
is the quadratic sum within each cluster. Here, $C$ is the test partition number in the index $C_i=\{C_1,C_2,\ldots, C_m\}$, $m$ is the number of clusters, $n_i$ is the number of elements in each $ith$ cluster, $n$ is the number of elements of the complete dataset, $x_j$ represents an element of the data set, $C_{P(j)}$ is the centroid associated with the partition $P(j)$ where the element $x_j$ is contained, and $\bar{x}$ is the mean value of the data within each cluster. According to \citet{Hartigan1975}, the optimum number of clusters is the smallest $k$ which produces $H(k)\leqslant \eta$ (typically $\eta=10$).

At the implementation of the $H(k)$ index, we iterate this method 50 times using its coded version in the R software package\footnote{R is a free software environment for statistical computing and graphics, available at \url{http://www.R-project.org/}.}. We detected an optimum number of $k = 16$ clusters to be applied to the sample, which satisfy the condition $H(k)\leqslant 10$. 
Once the optimum number of clusters is obtained, we estimate which data of a single ten-colour matrix belongs to each cluster by using the $k$-means.
It is important to say here that, KNN classifiers are used in two steps:

\begin{itemize}[leftmargin=0.7cm]
\renewcommand\labelitemi{--}
\item Initially by taking random centroids that are subsequently corrected by iterations, where in each iteration the centroid is recalculated in such a way that the distance between the centroids and the assigned objects is minimal. This technique is used to know which synthetic data belonged to each one of the 16 clusters.
\item After the cluster members are defined, fixed centroids for the observational data are used to estimate the closest synthetic dataset.
\end{itemize}

In particular, for the random centroid step, the iterations are performed with a convergence criterion of $0.001$ and the Euclidean distance is given by:
\begin{equation}
D_{ii'}=\sqrt{\sum_{j}\bigg(X_{ij}-X_{i'j}\bigg)^{2}},
\end{equation}
where $X$ is the value of each case $i$ and each centroid $i'$, for all the $j$ variables included in our study.

During the first step described above, we check the presence of observational and synthetic data in the 16 clusters. We found that the method generated a few outlier clusters composed by observational data only, comprising a total of 44 observed binary datasets. These binaries do not have similar characteristics to any synthetic colour, as they could not be compared to any model. Hence, they were discarded. After few iterations, the remaining 490 observational binary datasets go through the second step with fixed centroids. Finally, the temperatures are determined by the mean value from the synthetic data neighbouring each observational point and the formal errors are given by the standard deviation. 

The synthetic V+V models are represented in 8 of the 16 clusters generated by the KNN method. We found that 255 of our 534 DEB candidates present composite colours that would result from a combination of two main-sequence stars (V+V), which are in clusters with similarities to the V+V synthetic data. These objects are shown in Fig. \ref{obs_colourgrid}, 
which presents their distribution in some of the colour combinations obtained by the applied method, as an example. 
The colours represent the different clusters generated by the method, grouped by similarity. Later in the radius analysis, after modelling their light curves, 25 binaries among the selected V+V candidates were identified as having a giant star as component, since giants appear distant from the main-sequence distribution in a mass-radius diagram (see Sect. \ref{MRdiagram}). We then, selected 230 DEB candidates that correspond to systems composed by main-sequence stars (V+V) only. These objects are shown in Fig. \ref{NIR-CMD-534stars}, as filled circles.

We use the temperatures obtained for each component to determine the mass of both primary and secondary stars, via interpolation, following the tabulated semi-empirical values of stellar colours and effective temperature sequence by \citet{Pecaut12} and \citet{Pecaut13}\footnote{An updated version of the adopted table of stellar colours and effective temperatures is available at \url{http://www.pas.rochester.edu/~emamajek/EEM_dwarf_UBVIJHK_colours_Teff.txt}.}, which were obtained from a heterogeneous sample. The associated temperature uncertainties are considered for the estimation of the mass errors. In columns 2 to 5 of table \ref{Table:System-parameters} we present the effective temperatures and masses determined for the final sample of 230 DEB systems.

An important step during the development of this procedure was to test the reliability of the method described above using known control samples. For that, we have performed a simulation with synthetic binaries, where all components have known luminosity classes and temperatures. These binaries were generated with standard magnitudes from the available synthetic photometry from \citet{Bressan2012}. For each created binary, composite synthetic colours were estimated and a Gaussian noise was added in order to mimic observed magnitudes. The bootstrapping of the synthetic defined colours was performed using the typical standard deviation found in our observed sample. The created set of synthetic binaries, containing all three possible luminosity class combinations (V+V, V+III, and III+III), were submitted to our method with the objective of rejecting systems containing giants. The method  was found effective to eliminate undesired binaries, which is any system containing a giant star (V+III or III+III). From this set of synthetic binaries with giants, only 6.7\% of them have been wrongly selected as a V+V system by our method. It is worth to mention again that any giant contaminants would be found later in the analysis, since spurious radii can be identified in a mass-radius diagram. 
We also quantified that only 4.3\% of the V+V synthetic systems were discarded by the method, as false V+III or III+III binaries.

We also have checked the results for temperature and mass obtained by our method for a small control sample composed of low-mass binaries from the literature, with available 2MASS and SDSS colours. The components of the binaries in the control sample have well defined masses, which were derived spectroscopically. All masses and $T_{\rm eff}$ derived by our method, as well as the values from the literature, are shown in Table \ref{Table-Literature-LMB}, for the test control sample. The temperature uncertainties from our $k$-means method are calculated from the statistical uncertainty. Consistent temperatures were derived, which confirms the reliability of our method. The differences in masses can be understood since ours are derived from theoretical models and the values from the literature come from actual radial velocity mass functions.

\section{Light curve modelling with the JKTEBOP code and the AGA algorithm}\label{modelling}

All 230 light curves were modelled by using a combination of two codes: the JKTEBOP\footnote{JKTEBOP is written in FORTRAN 77 and the source code is available at \url{http://www.astro.keele.ac.uk/jkt/codes/jktebop.html}.} code \citep{Southworth2004} and a modified version of the asexual genetic algorithm \citep[AGA,][]{Canto2009}. The AGA algorithm was implemented as an additional task in the JKTEBOP code by \citet{Coughlin2011}. We adopted this mentioned additional task for its simplicity, speed, low numerical noise and reliability on solving similar problems, since it consists in the search for a global minimum over a large and potentially discontinuous parameter space. All details about the implementation and the specific changes in the original JKTEBOP code are described in \citet[][Appendix B]{Coughlin2011}.

The light curve modelling was performed to estimate the orbital and physical parameters of all DEB candidates in our sample. For that, we adopted the orbital period $P_{0}$ and epoch $T_{0}$ from Sect. \ref{select} as initial conditions and allowed the following parameters to vary in the fitting:

\begin{enumerate}
\addtolength{\itemindent}{0.7cm}
\item $J=J_{2}/J_{1}$, the central surface brightness ratio,
\item $(R_{1}+R_{2})/a$, the sum of the stellar radii, in units of the binary separation,
\item $k=R_{2}/R_{1}$, the ratio of the radii,
\item $i$, the orbital inclination,
\item $ecc$, the eccentricity of the orbit,
\item $\omega$, the argument of periastron,
\item $P_{0}$, the binary orbital period,
\item $T_{0}$, the reference time of primary minimum, and
\item the baseline level of the light curve.
\end{enumerate}
Given the large number of parameters in the model, and due to the limited precision of the photometric data, we have fixed some of the model parameters according to the following assumptions. 

We set both limb-darkening (LD) coefficients to fixed values in the fitting process. The LD coefficient is dependent on the stellar temperature, so, we have adopted the temperatures obtained in section \ref{Spectral-types} for each component. We used the JKTLD\footnote{JKTLD is written in FORTRAN 77 and the source code is available at \url{http://www.astro.keele.ac.uk/jkt/codes/jktld.html}} procedure to obtain the coefficients from \citet{Claret2000}, using a quadratic LD law. 
We also fixed the gravity darkening coefficients to the typical value adopted for stars with convective envelopes \citep[$\beta=0.32$, ][]{Lucy1967}. 

To estimate the photometric mass ratio ($q$), we adopted the power-law relation of $q\sim (R_{2}/R_{1})^{1.534}$ \citep{Svechnikov1983}, as suggested by \citet{Devor2008}. It is important to emphasise that the JKTEBOP code uses this mass ratio only to determine the amount of tidal deformation of the components, which is expected to be negligible for the present analysis. Thus, this parameter was kept fixed in the minimisation. Also, the reflection coefficients were set to fixed values, estimated by the code based on the system geometry.

A few tests were performed to fit for a third light ($l_{3}$). The obtained results were consistent with $l_{3}=0$. Moreover, we also have found that leaving $l_{3}$ free to vary did not improve significantly the fit. Therefore, $l_{3}$ was fixed to zero.

Besides the evaluation of the parameter uncertainties, all light curves were visually inspected to avoid including those with obvious discrepancies between the best-fitting model and the observations. Then, to obtain a robust error estimation for the fitted parameters, we performed Markov chain Monte Carlo (MCMC) simulations, which is already implemented in the JKTEBOP code, with 10000 iterations for each light curve. All obtained results are presented in table \ref{Table:System-parameters}\footnote{The complete version of this table is available online.}.

\begin{figure}
    \centering
\includegraphics[width=0.47\textwidth, angle=0]{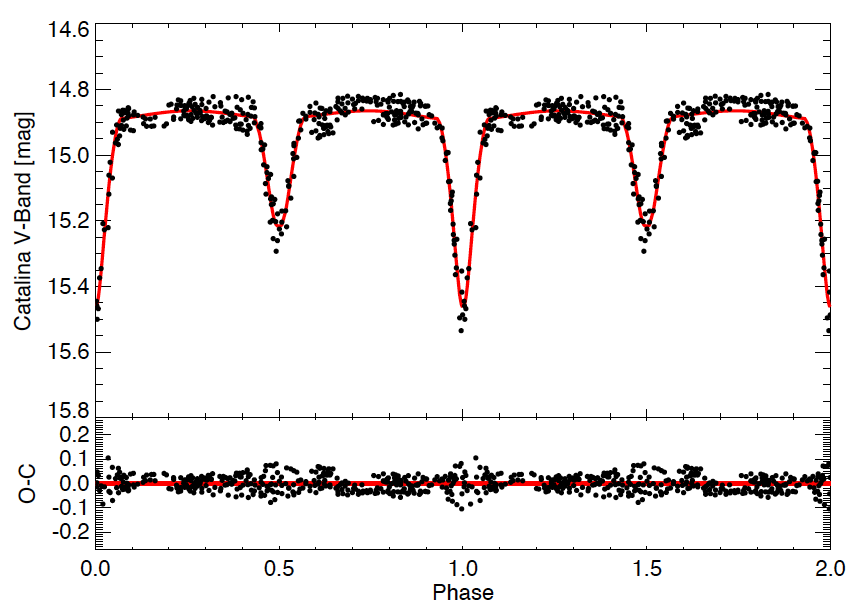}   
    \caption{Phase-folded light curve of CSSJ034302.8+0.109354 (Catalina ID: 1101020024323). The best fitting model is represented by a solid line. The lower panel shows the residuals of the fit.}
\label{Model}
\end{figure}

As an example, the phase-folded light curve of the system CSSJ034302.8+0.109354 (Catalina ID: 1101020024323) is presented in figure \ref{Model} along with the best-fitting model, which is represented by a solid line. The primary eclipse minimum corresponds to phase zero.

 \section{Discussion}\label{discuss}
 
We estimated individual masses from the method described in Sect. \ref{Spectral-types}. Additionally, we derived the fractional radius ($R/a$) for both components and the orbital period from the light curve model, described in Sect. \ref{modelling}. Following Kepler's third law, we estimated the orbital separation and hence the individual radius for the complete sample. The derived radii are presented in table \ref{Table:System-parameters}.

\subsection{The mass-radius diagram}\label{MRdiagram}

  \begin{figure*}
    \includegraphics[width=0.495\textwidth]{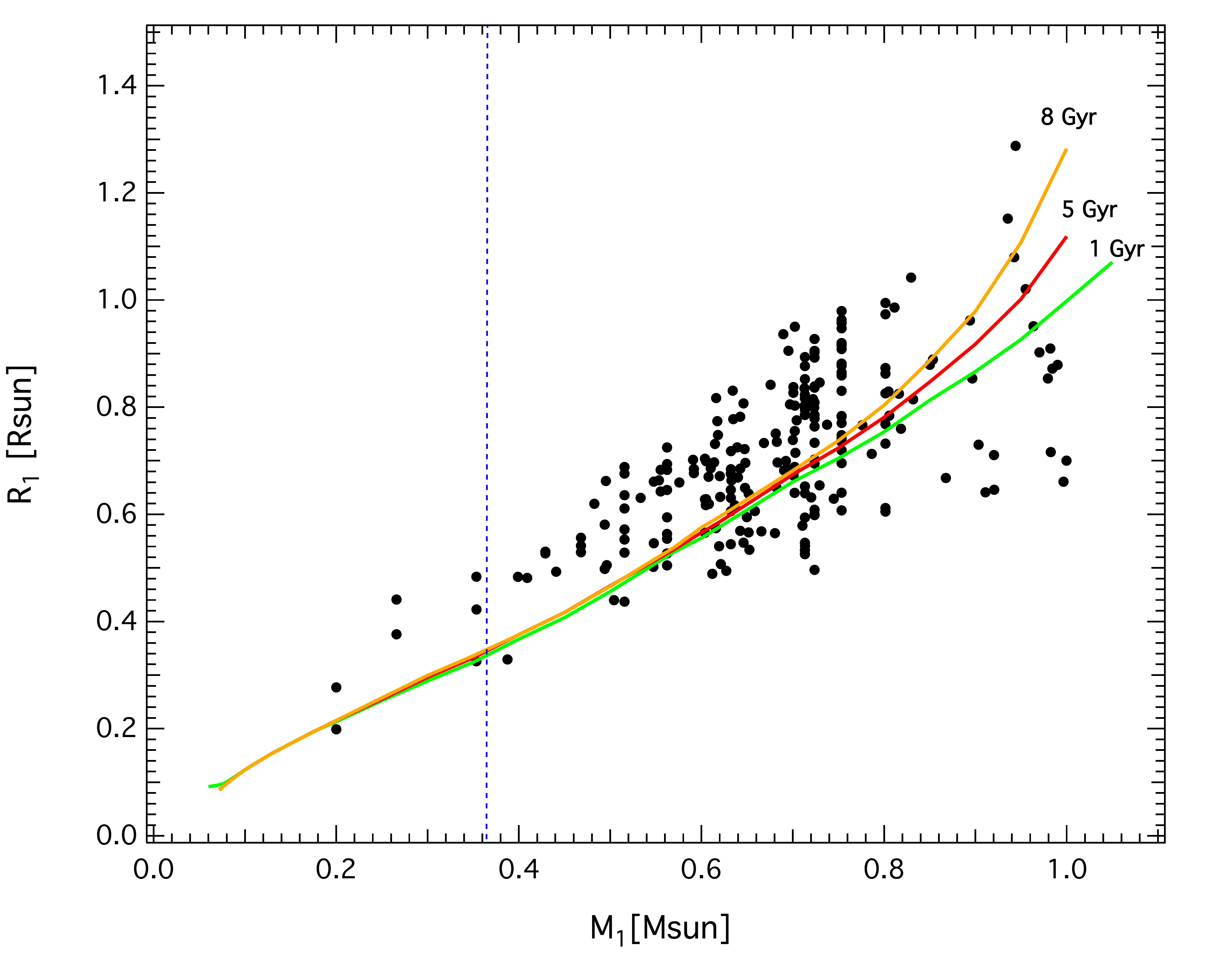}  
        \includegraphics[width=0.495\textwidth]{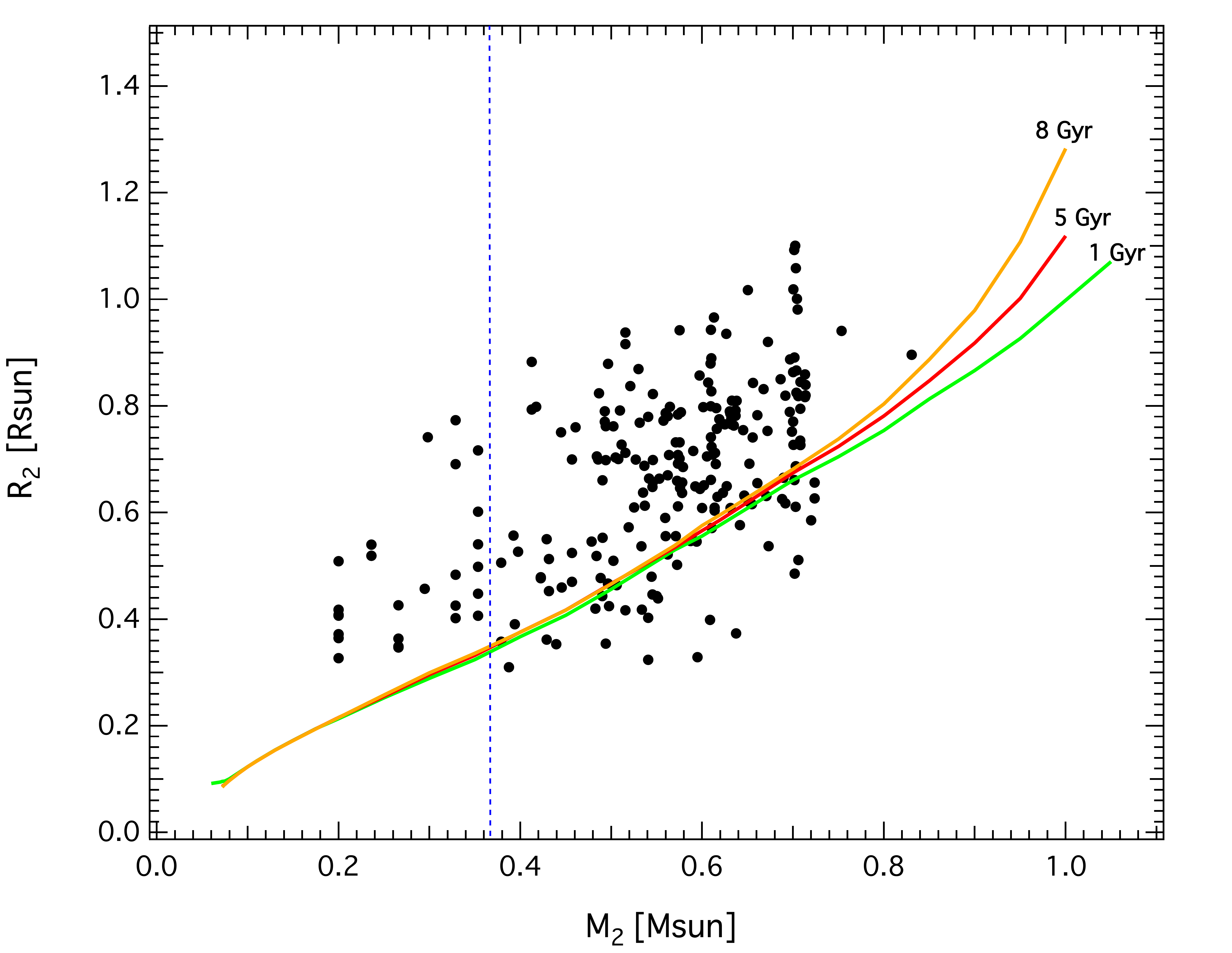}  
    \caption{Mass-radius diagram for 230 DEBs, with component masses of less than $1.0$ $M\odot$ and orbital periods shorter than 2 days. The left panel shows the diagram for the primary components and the right panel for the secondaries, individually. The theoretical models from \citet{Baraffe1998} for $1$, $5$, and $8$ Gyr are also shown, which are represented by green, red, and yellow solid lines, respectively. The vertical dotted line in both panels marks the limit between fully convective and partially radiative stars \citep[0.35 M$\odot$,][]{Chabrier1997}.}
 \label{Mass-Radius}
\end{figure*}

\begin{table*}
\centering
\begin{tabular}{c c c c c c c}
\hline
    & Mass range & \multicolumn{2}{c}{Number of} & Number of stars & Percentage & Statistical \\
 & (M$_{\odot}$) & Primaries & Secondaries & per bin & ($\%$) & significance \\\hline
{\footnotesize BIN 1} &  0.70 < M < 1.00 & 118 & 35 & 153 & 33.26\,\% & 0.655 \\
{\footnotesize BIN 2} &  0.56 < M < 0.70 & 66 & 87 & 153  & 33.26\,\% & 0.005 \\
{\footnotesize BIN 3} &  0.20 < M < 0.56 & 46 & 108 & 154  & 33.48\,\% & 0.005 \\
\hline\end{tabular}
\caption{Distribution of bins of mass for the DEBs-LMMS stars in our sample, with a total of $230$ primaries and $230$ secondaries. Each bin is composed approximately by a third of the whole sample ($\sim$$33\%$). The last column shows the statistical significance obtained from the KS test, described in Sect. \ref{MRdiagram}.}
\label{Table-bins}
\end{table*}

\begin{figure}
    \includegraphics[width=0.99\columnwidth]{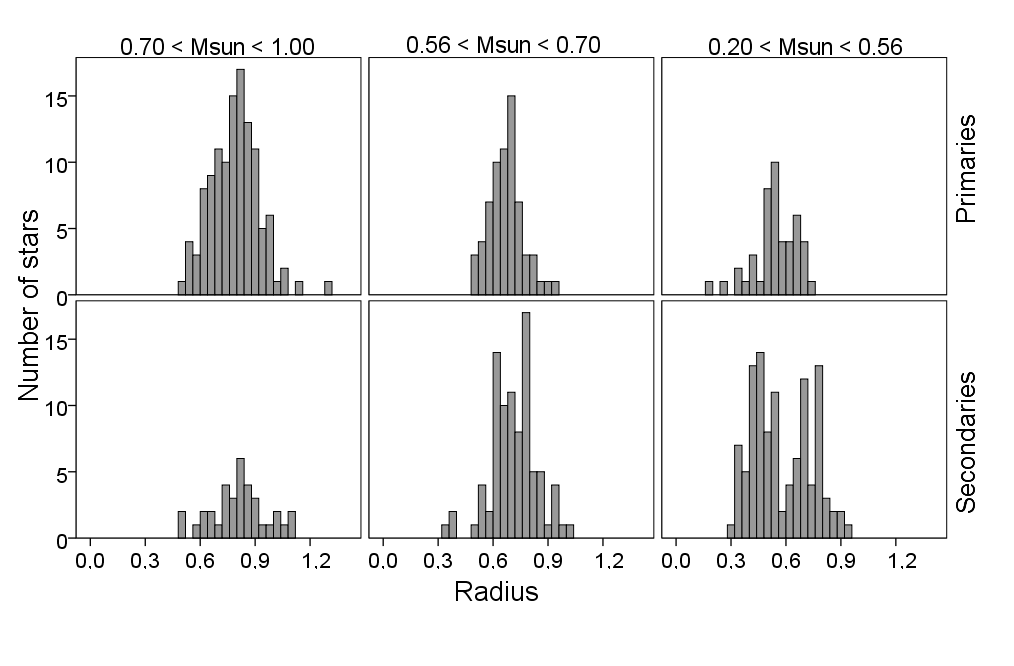}
    \caption{Histogram of distributions of radii in each bin of mass, where {\scriptsize BIN 1} comprehends masses from $0.70$ to $1.00$ M$_{\odot}$, {\scriptsize BIN 2} goes from $0.56$ to $0.70$ M$_{\odot}$, and {\scriptsize BIN 3} from $0.20$ to $0.56$ M$_{\odot}$ (see table \ref{Table-bins}). The upper panels show the distributions for primaries and the lower panels for secondaries.}
\label{Histogram}
\end{figure}

We focused this discussion to one of the current issues involving close-orbiting low-mass DEBs: the radius anomaly in the mass-radius diagram. The radius inflation problem is a well known issue in close-orbiting low-mass systems, especially in binary components with masses of less than $\sim$$0.6$ M$\odot$.

Figure \ref{Mass-Radius} presents the mass-radius diagram of the investigated sample, showing the distribution for primary and secondary components separately, on the left and right panels, respectively. Here, we also illustrate isochrones of $1$, $5$, and $8$ Gyr, which are represented by green, red, and yellow solid lines, respectively. These are theoretical models from \citet{Baraffe1998}, computed with a convective mixing length equal to the scale height, a helium abundance of $Y=0.275$, and with solar metallicity.

The radius inflation is conspicuous in our sample, clearly increasing towards lower masses. Despite the expected scatter present in the diagram (fig. \ref{Mass-Radius}) due to the limited precision of the analysed light curves and the uncertainties in the mass estimate procedure, it is possible to notice that the distributions of primaries and secondaries in the diagrams show a distinct behaviour. It is visible that the scatter present on the right panel of fig. \ref{Mass-Radius}, which illustrates only the secondary components, is greater than the scatter present on the left panel, for primaries. The average radius for low-mass secondaries is also larger than those for primaries of same mass. This, in fact, suggests that the sample of secondaries are more inflated than the sample of primary components.

To confirm what is inferred from fig. \ref{Mass-Radius} --that secondary components seems to be more inflated than primaries in a low-mass regime--, we have applied the two-sample Kolmogorov-Smirnov (KS) test. The KS test was used to verify if these two populations of stars (primaries and secondaries) come from the same distribution, i.e., if they follow the same global behaviour. To perform such analysis, we divided the whole sample in three bins of different mass intervals, from the more massive to the less massive components: from $0.70$ to $1.00$ ({\footnotesize BIN 1}), from $0.56$ to $0.70$ ({\footnotesize BIN 2}), and from $0.20$ to $0.56$ ({\footnotesize BIN 3}) M$_{\odot}$. 
These selected ranges of mass were defined to have approximately a third of the whole sample in each bin, as shown in table \ref{Table-bins}. We separated the sample in bins of mass in order to analyse the behaviour of the stellar radius in different regions of the low main-sequence stars, going from the most to the less massive objects in the sample. From the obtained distribution of radii, we can investigate if the radius inflation equally affects both components of the binary.

Figure \ref{Histogram} presents the histograms generated in order to get a visual impression of the radius distribution within each bin. Part of the spread in those distributions is due to the mass bin width and the slope of the mass-radius relation. In this figure, the three upper panels show the distributions obtained for primaries and the three lower panels for the secondary components. If we compare both panels on the left ({\footnotesize BIN 1}), comprehending the most massive objects ($0.70 < M < 1.0$ M$_{\odot}$), we are not able to reach any conclusions since the population of secondary components within this mass range is small --which was expected since secondaries are in general less massive than primaries and our sample is limited to low-mass objects. analysing the middle panels ({\footnotesize BIN 2}; $0.56 < M < 0.70$ M$_{\odot}$), both distributions may seem similar, however, the secondary component distribution present a wider range of radii with a suggestion of a bimodal distribution.

The most discrepant result is found for panels on the right of Fig. \ref{Histogram} ({\footnotesize BIN 3}; $0.20 < M < 0.56$ M$_{\odot}$), which is composed by the less massive objects in our sample, where the difference between primaries and secondaries become significant. Here, the distribution obtained for the secondary components (fig. \ref{Histogram}, right lower panel) clearly reaches larger radii than those obtained for primaries (fig. \ref{Histogram}, right upper panel) for the same mass range, and also shows a bimodal behaviour. At this point, it becomes more evident that the trend for more inflated secondary components shown in fig. \ref{Mass-Radius} is a real trend.

The histograms in figure \ref{Histogram} visually present the results obtained in the performed KS test. They suggest that only the more massive stars in our sample ({\footnotesize BIN 1}) could have come from the same distribution and follow the same global behaviour. This interpretation is supported by the KS test, where a high statistical significance, in a range from 0 to 1, would mean that the hypothesis of having both primary and secondary star samples drawn from the same population is probably true. The most massive stars in our sample, with a significance of 0.655, is the regime where we probably have primaries and secondaries with the same behaviour. The statistical significance given by the KS test is presented in table \ref{Table-bins}.

For the less massive objects in our sample ($M<0.70$ $M_{\rm \odot}$), the significance of the KS test is of only 0.005. This result rejects the tested hypothesis and tells us that, probably, the primaries and secondaries in the analysed sample present different distributions, with different behaviours. In other words, the KS test confirms that the secondary components in our sample are, in fact, more inflated than the primaries.

\subsection{The radius inflation of secondary components in the literature}

Recently, \citet{Cruz2018} have gathered a list of low-mass detached binaries from the literature composed by main-sequence stars with masses and radii of $0.7$ $M_{\odot}$ and $0.7$ $R_{\odot}$ or less, respectively. These objects have small uncertainties on the derived masses and radii, of less than $6\%$.

Comparing the compilation from \citet[][Appendix A]{Cruz2018} to the evolutionary model for $5$ Gyr from \citet{Baraffe1998}, we have found that half of the stars presented in their list have radii that are more than $5\%$ bigger than expected. We also have found that $14$ from the $23$ EB systems have secondaries more inflated than primary components, which comprises $61\%$ of the well-characterised DEBs from the literature \citep[][and references therein]{Cruz2018}. 

These authors also have characterised five low-mass EBs using radial velocity masses. Four of them also present more inflated secondary components. For instance, the secondary component of the EB 17e-3-02003, with mass of $M_2=0.51$ $M_{\odot}$, is $12.5\%$ inflated with respect to the $5$ Gyr model, however, the primary component, with mass of $M_1=0.60$ $M_{\odot}$, is only $8.6\%$ inflated, approximately \citep{Cruz2018}. 
There are yet other individual examples, like the EB MG1-116309 \citep{Kraus2011} with inflation of only $\sim$$3.8$ and $\sim$$6.9\%$ for the primary ($M_1=0.57$ $M_{\odot}$) and secondary ($M_2=0.53$ $M_{\odot}$) components, respectively, or the EB HATS557-027 \citep{Zhou2015} with an inflation of $4.3\%$ for the primary ($M_1=0.24$ $M_{\odot}$) and of $10.2\%$ for the secondary ($M_2=0.18$ $M_{\odot}$) component, approximately.

However, there are some systems that do not have more inflated secondaries. For example, the EB GJ 3236 \citep[$M_1=0.38$ $M_{\odot}$, $M_2=0.28$ $M_{\odot}$;][]{Irwin2009} is a short-period low-mass system where both components present a similar inflation of around $\sim$$7\%$. There are also systems that follow the evolutionary models and do not show any appreciable inflation, like the EB SDSS-MEB-1 \citep[$M_1=0.27$ $M_{\odot}$, $M_2=0.24$ $M_{\odot}$;][]{Blake2008}, which primary and secondary components present a negligible inflation of only $\sim$$1.8$ and $\sim$$0.4\%$, respectively.

The radius inflation problem of low-mass stars as components of close-orbiting EB systems seems to be significant for at least half of the well-characterised DEBs in the literature. This work suggested that, despite the limitation on precision, the investigated sample follow an inflation trend, especially in secondary components with masses of less than $\sim$$0.6$ M$\odot$. We emphasise the importance of increasing the sample of known short-period DEB systems, with homogeneously derived masses and radii, to correlate variables and investigate the causes of radius inflation. Such study would help to improve theoretical stellar structure and evolutionary models.

\section{Conclusions}\label{concl}

This work was dedicated to identify and photometrically characterise a sample of detached close-orbiting low-mass eclipsing binary systems in the Catalina Sky Survey. The identification and candidate selection was carefully performed through several steps to compose a sample of detached systems with orbital periods shorter than $2$ days, with main-sequence components. 
A SDSS-2MASS ten-colour grid of composite synthetic and observed colours and the K-Nearest Neighbours method were employed to effectively define a final sample of 230 DEB candidates with estimated effective temperatures of $T_{\rm eff}\leq 5720$ K and masses of $M \leq 1.0\ M_{\odot}$. All light curves were modelled by using a combination of the JKTEBOP code and an asexual genetic algorithm yielding orbital parameters such as the binary orbital period and inclination and the fractional radius of the components. The adopted approach provided the opportunity to estimate basic stellar parameters (as mass and radius) in large databases, by using only light curves and multi-band photometry.

Our sample with a statistically significant number of short-period low-mass binary systems allowed the study of the distribution of inflation in the mass-radius diagram. Inflation increases towards lower masses and our results suggest that the secondary components are more inflated than primary components of the same mass regime in low-mass DEB systems.

Comparing other low-mass EBs from the literature to an evolutionary model of $5$ Gyr, we have found that around $61\%$ of the systems also have secondaries more inflated than primaries. Interestingly, we also noted that half of the components of these systems present radius $5\%$ or more larger than expected from models.

The radius inflation problem for stars with masses of less than $\sim$$0.6$ M$\odot$ has been around for more than 15 years. It affects several areas of stellar astrophysics where the mass-radius calibration plays an important role. Therefore, it is important to increase the sample of DEB systems with derived masses and radii to better understand the causes of such inflation and to improve the present theoretical models of low-mass stars.

\section*{Acknowledgements}

The authors acknowledge the referee for all suggestions and comments that have considerably improved the present paper. We also acknowledge Dr. Jeffrey Coughlin for providing the modified version of the JKTEBOP code, used in the work. This research has been funded by Brazilian CNPq grant 152237/2016-0 and FAPESP grant 2015/18496-8. MD thanks CNPq funding under grant \#305657. This study was financed in part by the Coordena\c c\~ao de Aperfei\c coamento de Pessoal de N\'ivel Superior - Brasil (CAPES) - Finance Code 001. This publication makes use of data products from the Two Micron All Sky Survey, which is a joint project of the University of Massachusetts and the Infrared Processing and Analysis Center/California Institute of Technology, funded by the National Aeronautics and Space Administration and the National Science Foundation.






\appendix 
\section{Derived parameters for the complete sample of 230 DEB systems.}

We present in table \ref{Table:System-parameters} all derived parameters for our complete sample of 230 detached EBs found in the Catalina Sky Survey. The identification Catalina ID number is presented in column 1. The derived effective temperatures and masses, shown in columns 2 to 5, were obtained from a grid of composite synthetic and observed colours, as described in detail in Sect. \ref{Spectral-types}. 
The quantities presented in columns 6 to 13, e.g. orbital parameters and fractional radii, were obtained from the light curve modelling by using a combination of the JKTEBOP code \citep{Southworth2004} and the AGA algorithm \citep{Canto2009}, as performed by \citet{Coughlin2011}, as described in Sect. \ref{modelling}. The last two columns, 14 and 15, show the radii obtained by using Kepler's third law.  A complete version of this table is available online.
\\

\begin{landscape}
\renewcommand{\arraystretch}{1.0}
{\setlength{\doublerulesep}{5mm}
\begin{table}\caption{Orbital and Physical parameters for the 230 DEBs with LMMS stars found in the Catalina Sky Survey, with mass of $ M\leq 1.0\, M_{\odot}$ and orbital period shorter 2 days. The values of $e\cos \omega$ and $e\sin \omega$ were solved  in the analysis for present the values for $e$ and $\omega$ in this table. (This table is fully published online.)}
\label{Table:System-parameters}
\begin{small}
\scalebox{0.85}[0.65]{
\begin{tabular}{|r|r|r|r|r|r|r|r|r|r|r|r|r|r|r|}
\hline
  \multicolumn{1}{|c|}{Catalina ID} &
  \multicolumn{1}{c|}{T$_{1}$} &
  \multicolumn{1}{c|}{T$_{2}$} &
  \multicolumn{1}{c|}{M$_{1}$} &
  \multicolumn{1}{c|}{M$_{2}$} &
  \multicolumn{1}{c|}{J} &
  \multicolumn{1}{c|}{$i$} &
  \multicolumn{1}{c|}{Period} &
  \multicolumn{1}{c|}{T$_{0}$} &
  \multicolumn{1}{c|}{$R_1/a$} &
  \multicolumn{1}{c|}{$R_2/a$} &
  \multicolumn{1}{c|}{$e$} &
    \multicolumn{1}{c|}{$\omega$} &
  \multicolumn{1}{c|}{R$_{1}$} &
  \multicolumn{1}{c|}{R$_{2}$} \\
    \multicolumn{1}{|c|}{} &
  \multicolumn{1}{c|}{(K)} &
  \multicolumn{1}{c|}{(K)} &
  \multicolumn{1}{c|}{(M$_{\odot}$)} &
  \multicolumn{1}{c|}{(M$_{\odot}$)} &
  \multicolumn{1}{c|}{} &
   \multicolumn{1}{c|}{($^{\circ}$)} &
  \multicolumn{1}{c|}{(Days)} &
  \multicolumn{1}{c|}{(MJD-2450000)} &
  \multicolumn{1}{c|}{} &
  \multicolumn{1}{c|}{} &
  \multicolumn{1}{c|}{($^{\circ}$)} &
   \multicolumn{1}{c|}{($^{\circ}$)} &
  \multicolumn{1}{c|}{(R$_{\odot}$)} &
  \multicolumn{1}{c|}{(R$_{\odot}$)} \\
\hline
		1001007032234	&	$	4222\pm	100$	&	$	3854\pm	162$	&	$	0.702\pm0.087$	&	$	0.601\pm0.127$	&	$	0.634	\pm	0.009	$	&	$	81.1	\pm	0.1	$	&	0.554433855	&	53627.29506	&	$	0.22	\pm	0.003	$	&	$	0.257	\pm	0.004	$	&	$	0.067	\pm	0.001	$	&	$	90.5	\pm	1.2	$	&	0.682	&	0.798	\\
	1001028026235	&	$	3832\pm	100$	&	$3569	\pm	118$	&	$0.592\pm0.119$	&	$0.484\pm0.013$	&	$	0.546	\pm	0.007	$	&	$	87.1	\pm	0.1	$	&	0.546642796	&	54829.2719	&	$	0.235	\pm	0.003	$	&	$	0.18	\pm	0.003	$	&	$	0.027	\pm	0.001	$	&	$	91.5	\pm	3	$	&	0.677	&	0.519	\\
	1001047013702	&	$	4366\pm	100$	&	$3993\pm	100$	&	$0.713\pm0.105$	&	$0.633\pm0.110$	&	$	0.837	\pm	0.011	$	&	$	81	\pm	0.1	$	&	0.420363803	&	54539.20921	&	$	0.202	\pm	0.003	$	&	$	0.31	\pm	0.004	$	&	$	0.016	\pm	0.001	$	&	$	75.1	\pm	4.6	$	&	0.526	&	0.810	\\
	1001047059600	&	$	4653	\pm	100$	&	$	4373\pm112	$	&	$0.738\pm0.137$	&	$0.714\pm0.107$	&	$	0.846	\pm	0.012	$	&	$	82.4	\pm	0.1	$	&	0.730598537	&	54573.14171	&	$	0.198	\pm	0.003	$	&	$	0.217	\pm	0.003	$	&	$	0.029	\pm	0.001	$	&	$	96.4	\pm	2.7	$	&	0.767	&	0.839	\\
	1001052064606	&	$	3861\pm	100$	&	$	3563\pm	100$	&	$0.604\pm0.113$	&	$	0.483\pm0.022$	&	$	0.744	\pm	0.01	$	&	$	84.3	\pm	0.1	$	&	0.280718196	&	53858.189	&	$	0.339	\pm	0.005	$	&	$	0.226	\pm	0.003	$	&	$	0.017	\pm	0.001	$	&	$	268.4	\pm	4.5	$	&	0.628	&	0.420	\\
	1001117004993	&	$	3707\pm	100$	&	$3601\pm 100$	&	$	0.533\pm0.014$	&	$	0.490\pm0.014$	&	$	0.752	\pm	0.01	$	&	$	85.2	\pm	0.1	$	&	0.660730229	&	55476.13263	&	$	0.196	\pm	0.003	$	&	$	0.138	\pm	0.002	$	&	$	0.042	\pm	0.001	$	&	$	272.4	\pm	1.9	$	&	0.631	&	0.443	\\
	1001123051042	&	$	4079\pm	100$	&	$	3703\pm	252$	&	$0.652\pm0.108$	&	$0.531\pm0.047$	&	$	0.665	\pm	0.009	$	&	$	87.2	\pm	0.1	$	&	0.552621817	&	55368.40994	&	$	0.189	\pm	0.003	$	&	$	0.256	\pm	0.004	$	&	$	0.052	\pm	0.001	$	&	$	92.5	\pm	1.5	$	&	0.566	&	0.769	\\
	1004010030420	&	$4207\pm100$	&	$	3791\pm	323$	&	$0.701\pm0.085$	&	$0.572\pm0.070$	&	$	0.3	\pm	0.004	$	&	$	86.9	\pm	0.1	$	&	0.495190478	&	54030.32391	&	$	0.29	\pm	0.004	$	&	$	0.176	\pm	0.003	$	&	$	0.086	\pm	0.001	$	&	$	281.1	\pm	0.9	$	&	0.827	&	0.502	\\
	1004029033156	&	$	4607\pm104	$	&	$	3894\pm	474$	&	$0.729\pm0.138$	&	$	0.613\pm0.185$	&	$	0.296	\pm	0.004	$	&	$	77.9	\pm	0.1	$	&	0.625289272	&	55602.1049	&	$	0.193	\pm	0.003	$	&	$	0.284	\pm	0.004	$	&	$	0.062	\pm	0.001	$	&	$	90	\pm	1.3	$	&	0.654	&	0.966	\\
	1004115044350	&	$4200\pm	100$	&	$	3878\pm	100$	&	$	0.700\pm0.084$	&	$	0.610\pm0.110$	&	$	0.639	\pm	0.009	$	&	$	78.5	\pm	0.1	$	&	0.872467504	&	53562.31535	&	$	0.176	\pm	0.002	$	&	$	0.19	\pm	0.003	$	&	$	0.006	\pm	0.001	$	&	$	123	\pm	12.5	$	&	0.739	&	0.799	\\
	1004115069102	&	$	4514\pm100	$	&	$3881\pm100	$	&	$0.724\pm0.123	$	&	$	0.610\pm0.111$	&	$	0.738	\pm	0.01	$	&	$	89.8	\pm	0.1	$	&	0.484935414	&	55478.17188	&	$	0.256	\pm	0.004	$	&	$	0.289	\pm	0.004	$	&	$	0.013	\pm	0.001	$	&	$	276.5	\pm	6	$	&	0.734	&	0.827	\\
	1004120029786	&	$	5632\pm186	$	&	$	4273\pm677	$	&	$	0.982\pm0.018$	&	$	0.706\pm0.208$	&	$	0.238	\pm	0.003	$	&	$	67.9	\pm	0.1	$	&	0.426867507	&	53937.36507	&	$	0.32	\pm	0.004	$	&	$	0.288	\pm	0.004	$	&	$	0.087	\pm	0.001	$	&	$	78.7	\pm	0.9	$	&	0.909	&	0.818	\\
	1004120047050	&	$	3769\pm100	$	&	$	3499\pm	204$	&	$0.562	\pm	0.136$	&	$	0.440\pm0.037$	&	$	0.358	\pm	0.005	$	&	$	86.4	\pm	0.1	$	&	0.551388801	&	55735.42829	&	$	0.245	\pm	0.003	$	&	$	0.125	\pm	0.002	$	&	$	0.021	\pm	0.001	$	&	$	100.7	\pm	3.7	$	&	0.694	&	0.353	\\
	1007112014719	&	$	3478\pm	100$	&	$	3357\pm	100$	&	$0.429\pm0.002	$	&	$0.353\pm0.026	$	&	$	0.839	\pm	0.012	$	&	$	86.2	\pm	0.1	$	&	0.854307228	&	55094.26951	&	$	0.152	\pm	0.002	$	&	$	0.116	\pm	0.002	$	&	$	0.005	\pm	0.001	$	&	$	3.1	\pm	\cdots	$	&	0.53	&	0.406	\\
	1007112081667	&	$	4366\pm	100$	&	$	4009\pm	146$	&	$0.713	\pm	0.105$	&	$0.635\pm0.120$	&	$	0.706	\pm	0.01	$	&	$	86.3	\pm	0.1	$	&	1.834443367	&	54765.1263	&	$	0.094	\pm	0.001	$	&	$	0.113	\pm	0.002	$	&	$	0.01	\pm	0.001	$	&	$	97.2	\pm	8.1	$	&	0.652	&	0.785	\\
	1007114053778	&	$	4723\pm	100$	&	$	4166\pm	109$	&	$0.753\pm0.136$	&	$0.686\pm0.093$	&	$	0.94	\pm	0.013	$	&	$	87.5	\pm	0.1	$	&	0.794232205	&	56076.40178	&	$	0.24	\pm	0.003	$	&	$	0.209	\pm	0.003	$	&	$	0.09	\pm	0.001	$	&	$	270.5	\pm	0.9	$	&	0.979	&	0.850	\\
	1007119012185	&	$	3927\pm	100$	&	$	3150\pm100	$	&	$	0.620\pm0.110$	&	$	0.200\pm0.096$	&	$	0.796	\pm	0.011	$	&	$	73.5	\pm	0.1	$	&	0.401933704	&	56099.59288	&	$	0.295	\pm	0.004	$	&	$	0.237	\pm	0.003	$	&	$	0.037	\pm	0.001	$	&	$	291.4	\pm	2	$	&	0.633	&	0.509	\\
	1101019022829	&	$	3798\pm	168$	&	$	3548\pm250	$	&	$0.576\pm0.142$	&	$0.478\pm0.037$	&	$	1.006	\pm	0.014	$	&	$	84.3	\pm	0.1	$	&	0.329702553	&	53763.12899	&	$	0.323	\pm	0.004	$	&	$	0.267	\pm	0.004	$	&	$	0.019	\pm	0.001	$	&	$	95.6	\pm	4	$	&	0.659	&	0.546	\\
	1101020024323	&	$	4056\pm100	$	&	$	3357\pm	100$	&	$0.642\pm0.114$	&	$0.353\pm0.026$	&	$	0.66	\pm	0.009	$	&	$	85.5	\pm	0.1	$	&	0.611937615	&	53709.24299	&	$	0.258	\pm	0.004	$	&	$	0.178	\pm	0.003	$	&	$	0.002	\pm	0.001	$	&	$	104.1	\pm	33.5	$	&	0.782	&	0.540	\\
	1101048064378	&	$	3992\pm	100$	&	$	3715\pm	245$	&	$	0.633\pm0.110$	&	$0.537\pm0.043$	&	$	1.004	\pm	0.014	$	&	$	90	\pm	0.1	$	&	1.099392778	&	55988.16408	&	$	0.14	\pm	0.002	$	&	$	0.13	\pm	0.002	$	&	$	0.01	\pm	0.001	$	&	$	82.3	\pm	7.4	$	&	0.664	&	0.613	\\
	1101054010186	&	$	3880\pm	100$	&	$3615\pm240$	&	$0.610\pm0.110$	&	$	0.493\pm0.045$	&	$	0.473	\pm	0.006	$	&	$	77	\pm	0.1	$	&	0.764002222	&	53494.21215	&	$	0.189	\pm	0.003	$	&	$	0.212	\pm	0.003	$	&	$	0.091	\pm	0.001	$	&	$	273.3	\pm	0.9	$	&	0.686	&	0.770	\\
	1101055014974	&	$4514\pm100	$	&	$3905\pm203$	&	$	0.724\pm0.123$	&	$	0.616\pm 0.130$	&	$	0.628	\pm	0.009	$	&	$	87.8	\pm	0.1	$	&	1.439709168	&	54881.26344	&	$	0.084	\pm	0.001	$	&	$	0.135	\pm	0.002	$	&	$	0.072	\pm	0.001	$	&	$	86.7	\pm	1.1	$	&	0.496	&	0.796	\\
	1101083064030	&	$	3754\pm	100$	&	$	3483\pm186	$	&	$0.555\pm0.140$	&	$0.432\pm0.032$	&	$	0.586	\pm	0.008	$	&	$	82.9	\pm	0.1	$	&	0.422483497	&	55618.45449	&	$	0.289	\pm	0.004	$	&	$	0.217	\pm	0.003	$	&	$	0.087	\pm	0.001	$	&	$	92.8	\pm	0.9	$	&	0.683	&	0.513	\\
	...	&	...	&	...	&	...	&	...	&	...	&	...	&	...	&	...	&	...	&	...	&	...	&	...	&	...	&	...	\\
  \hline\end{tabular}}
\end{small}
\end{table}}
\end{landscape}








\bsp	
\label{lastpage}
\end{document}